\documentclass[a4paper]{jpconf}

\usepackage[english]{babel}
\usepackage{multicol}
\usepackage{graphicx}
\usepackage{amssymb}
\usepackage{epsfig}
\usepackage{enumerate}
\usepackage{pifont}
\usepackage{hyperref}
\usepackage{verbatim}
\usepackage{txfonts}
\usepackage{graphicx}
 \newcommand{\jpsi}{{\rm J}/\psi}

\begin{document}

\title[PHENIX first measurement of the $\jpsi$ elliptic flow in Au+Au at $\sqrt{s_{\rm NN}}=200$~GeV]{PHENIX first measurement of the $J/\psi$ elliptic flow parameter $v_2$ in Au+Au collisions at $\sqrt{s_{\rm NN}}=200$~GeV}

\author{Catherine Silvestre (for the PHENIX Collaboration)}

\address{Dapnia, CEA-Saclay\\F-91191, Gif-sur-Yvette, France}
\ead{silvestr@rcf.rhic.bnl.gov}

\begin{abstract}
 
Recent results indicate that the $J/\psi$ suppression pattern differs with rapidity showing a larger suppression at forward rapidity. $J/\psi$ suppression mechanisms based on energy density (such as color screening, interaction with co-movers, etc.) predict the opposite trend. On the other hand, it is expected that more $c\bar c$ pairs should be available to form quarkonia at mid-rapidity via recombination. Some models provide a way to differentiate $J/\psi$ production from initially produced $c\bar c$ pairs and final state recombination of uncorrelated pairs, via the rapidity and transverse momentum dependence of the elliptic flow ($v_2$). \\
During 2007 data taking at RHIC, a large sample of Au+Au collisions at $\sqrt{s_{NN}}$=200 GeV was collected. The statistics has been increased compared to previous 2004 data set, thus allowing a more precise measurement of the $J/\psi$ production at both mid and forward rapidity. Furthermore, the PHENIX experiment benefited from the addition of a new detector, which improves the reaction plane resolution and allows us to measure the $J/\psi$ $v_2$. Comparing this measurement to the positive D-mesons $v_2$ (through non-photonic electron decays) will help constraining the $J/\psi$ production mechanisms and getting a more precise picture of the proportion of $J/\psi$ coming from direct production or charm quark coalescence.\\
Details on how the $J/\psi$ $v_2$ is measured at mid-rapidity rapidities are presented. The $J/\psi$ $v_2$ as a function of transverse momentum are compared to existing models.
\end{abstract}

\section{Introduction}

The use of charm quarkonia produced in ultra-relativistic heavy-ion collisions as quark gluon plasma (QGP) probes has a rich tradition, starting with the seminal paper by Matsui and Satz~\cite{MatsuiSatz}, which predicted the suppression of $\jpsi$ production in a QGP. At RHIC, PHENIX has measured the $\jpsi$ production in A-A collisions at $\sqrt{s_{NN}}=200~GeV$. The suppression as a function of the number of participants is found to be similar to the one at lower energy at SPS~\cite{NA50} even though the energy in the center of mass differs by an order of magnitude, leading to different cold nuclear effects (CNM) as well as the melting of other excited charmonium states. On the other hand, the larger energy and charm quark densities created at RHIC may allow regeneration of the $\jpsi$ via recombination of uncorrelated charm quarks during the collision~\cite{thews,andronic,capella}. Moreover, PHENIX recently published results~\cite{ppg68} indicate that the nuclear modification factor $R_{AA}$ is smaller at forward rapidity than at mid-rapidity. This may come from the amount of CNM effects that may be different for each rapidity~\cite{matt}, or from coalescence of charm quarks which would enhance the $\jpsi$ production further at mid-rapidity.

Studying other observables, such as the azimuthal anisotropy of the produced particles might give additional information on how these particles are produced~\cite{Zi-weiLin}. In non-central heavy ion collisions, the distribution of the colliding matter is not isotropic around the beam direction. In the thermodynamic picture, the asymmetric distribution of initial energy density causes a pressure gradient which is larger in the shortest direction of the ellipsoidal medium.  Therefore, the momentum distribution of produced particles can also be anisotropic. PHENIX measured a positive elliptic flow  for heavy (c,b) quarks via the study of so called non photonic electrons for a transverse momentum higher than 1~GeV/c~\cite{Alan}. 
If $\jpsi$ are produced from $c \bar c$ recombination in a deconfined phase~\cite{Robert L. Thews}, they should inherit their flow. On the contrary, $\jpsi$ directly produced by hard QCD processes early in the collision cannot be sensitive to collective phenomena. Therefore measuring a positive $\jpsi$ elliptic flow would indicate the level of recombination that takes part in the $\jpsi$ production mechanisms~\cite{AndronicBraun-Munzinger}.
 
\section{$\jpsi$ elliptic flow}

The elliptic flow is quantified by the second Fourier coefficient $v_2$ of the $\jpsi$ azimuthal angle distribution~\cite{bib:Ollitrault2,bib:Poskanzer}:
\begin{equation}\label{v2 fourrier}
\frac{dN}{d(\phi-\psi)} = N_0[1+2\cdot v_2\cdot cos(2\cdot(\phi-\psi))]
\end{equation}
with $\phi$ the azimuthal angle of the particle with respect to an arbitrary axis (Oj), and $\psi$, the reaction plane angle (plane defined by the impact parameter vector and the beam axis) with respect to (Oj).

The reaction plane of each event is measured using the azimuthal distribution of a limited subset of particles produced in the collision. Its resolution depends on the number of particles detected, the strength of their flow, and the resolution of the detector. Therefore, the elliptic flow measurement is corrected so that this resolution is reflected on the true $v_2$ measurement: $v_2^{true} = v_2^{meas}/\sigma_{RP}$.\\

\begin{figure}[h]
\begin{minipage}{20pc}
\includegraphics[width=20pc]{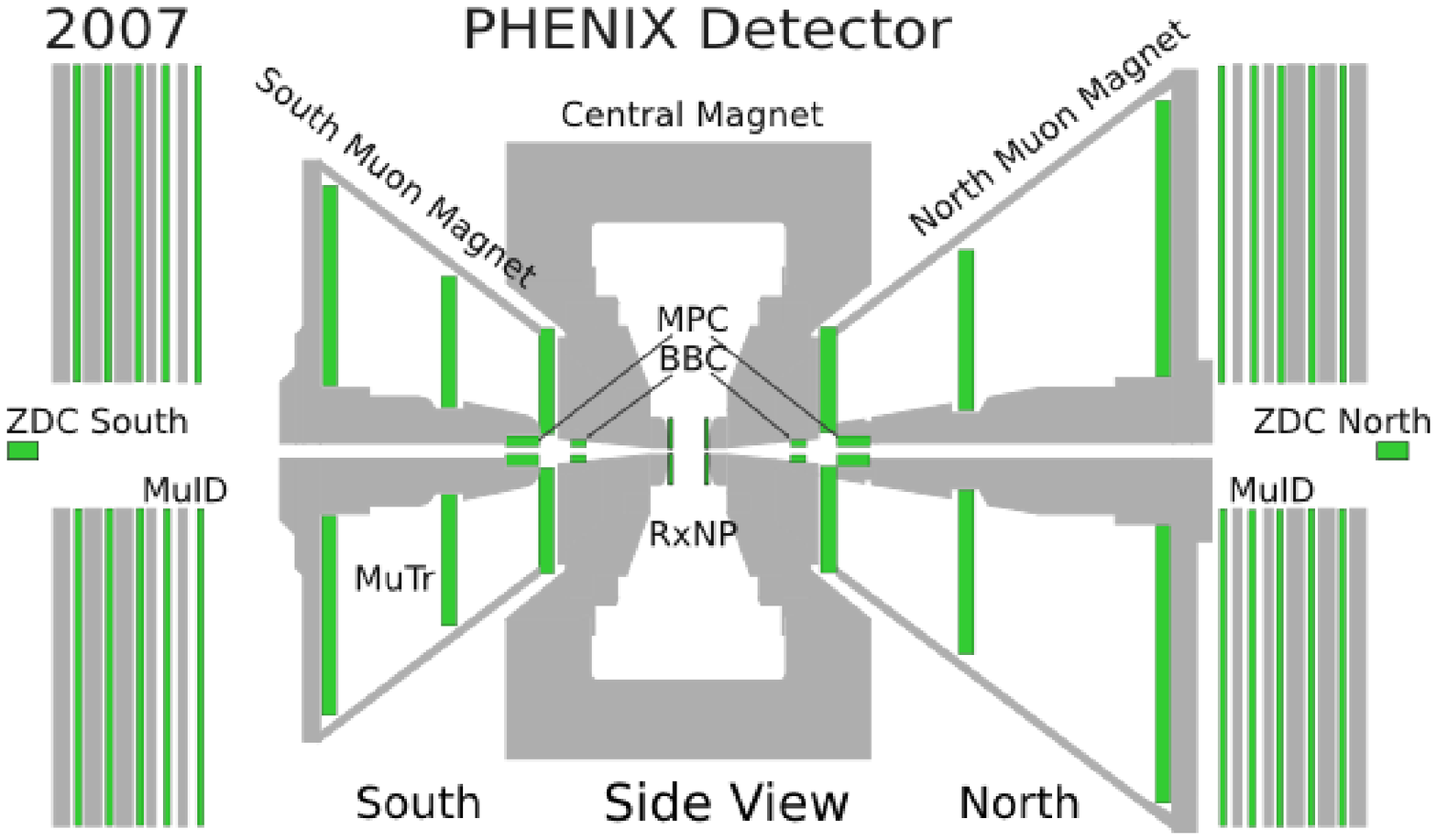}
\caption{\label{fig:rxnp_fig}PHENIX detector during 2007 data taking with the RxnP detector near the collision vertex.}
\end{minipage}\hspace{1pc}%
\begin{minipage}{17pc}
\includegraphics[width=15pc]{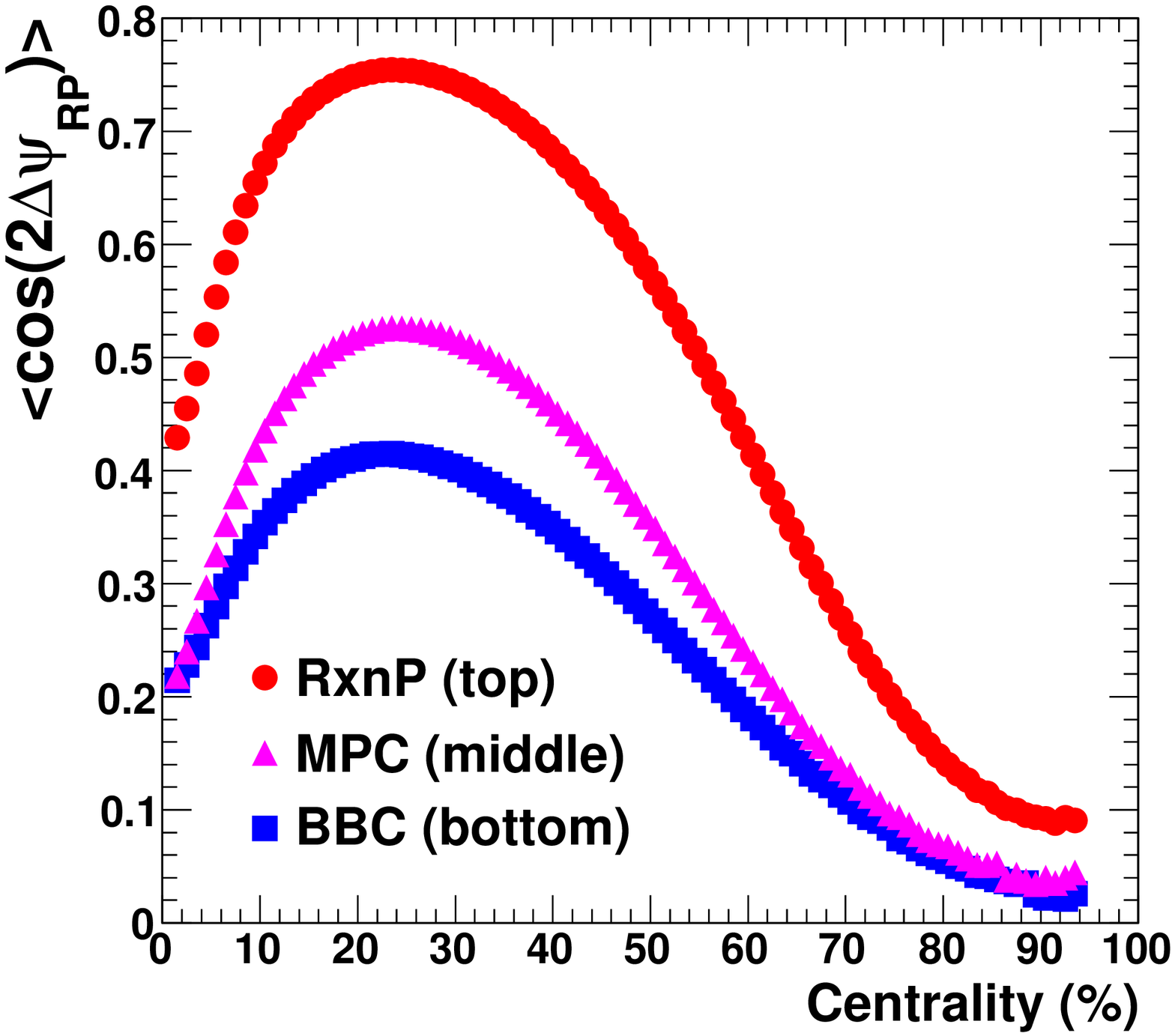}
\caption{\label{fig:rxnp}Reaction plane resolution as a function of centrality, measured with the RxnP (top), the MPC  (middle), or the BBC (squares).}
\end{minipage} 
\end{figure}

Starting from 2007, PHENIX uses two new Reaction Plane Detectors (RxnP) (see Fig.~\ref{fig:rxnp_fig}) to measure the reaction plane of each collision following methods described in~\cite{bib:Poskanzer,bib:Ollitrault}. It improves the reaction plane resolution, and thus gives a correction $\sigma_{RP}=\langle\cos(2\Delta\psi_{RP})\rangle$ twice better than what was achieved previously with the Beam Beam Counters (BBC), or what can be measured with the Muon Piston Calorimeters (MPC) (see Fig.~\ref{fig:rxnp}). 

The PHENIX experiment is designed to detect heavy quarkonia at forward rapidity ($1.2<|y|<2.2$) via the $\mu^+\mu^-$ decay channel and at mid-rapidity ($|y|<0.35$) via the $e^+e^-$ decay channel~\cite{phenix}. At forward rapidity, cathode strip tracking chambers and alternating layers of steel absorber and Iarocci tubes allow muons to be identified and tracked over an acceptance of $\Delta\phi= 360^\circ$. At mid-rapidity drift chambers, ring imaging \v{C}erenkov detectors and electromagnetic calorimeters are used to detect electrons in two arms covering $\Delta\phi=90^\circ$ each. During the 2007 run at the Relativistic Heavy Ion Collider, PHENIX recorded Au+Au collisions and $1.8\times10^9$ events are analysed here (42\% of the total luminosity). 

The forward measurement of $\jpsi$ elliptic flow benefits from three times more $\jpsi$ than at mid-rapidity but with a poorer resolution (150~MeV vs. 80~MeV) and signal over background ratio (1/6 vs. 1/3). In addition, at mid-rapidity both RxnP detectors can be used leading to an optimal measurement of $v_2$. A challenge to the forward rapidity measurement comes from the overlapping acceptances of the RxnP with the muon arms (see Fig.~\ref{fig:rxnp_fig}). The two muons and all possible accompanying fragmentation products might go through the RxnP, which may introduce non-flow effects and auto-correlations between the reaction plane and the $\jpsi$ azimuthal angle. To limit this bias on the $v_2$ measurement, the best configuration uses the RxnP located in the muon arm opposite to the $\jpsi$. The drawback is that the resolution of the reaction plane measurement is worse than when using both forward and backward RxnP detectors. On the other hand, using both RxnP or restricting to the one where the muons go (and thus possibly biasing the measurement) could be used to estimate the magnitude of the bias.

$\jpsi$ $v_2$ results are still underway at forward rapidity even though the $R_{AA}$ has been measured~\cite{Oda} and is compatible with published results~\cite{ppg68}. The method consists in extracting the number of $\jpsi$ counts in bins of $\phi$-$\psi$, transverse momentum $p_T$ and centrality using mixed event background subtraction. The $v_2^{meas}$ is extracted from the fit of the invariant yield with Eq.~\ref{v2 fourrier} in each $p_T$ and centrality bin, and is corrected with $\sigma_{RP}$. A similar method was also applied for mid-rapidity, but the result presented here uses an alternate method based on the additivity of the elliptic flow~\cite{olli_borgi}, which  provides a more precise handling of the background elliptic flow:
\begin{equation}
v_2^{S}(M_{\jpsi}) = v_2^{FG}(M_{\jpsi})\cdot\frac{N^{FG}(M_{\jpsi})}{N^{S}(M_{\jpsi})} - f^{BG}(M_{\jpsi}) \cdot\frac{N^{comb}(M_{\jpsi})}{N^{S}(M_{\jpsi})}
\end{equation} 
$v_2^{FG}(M_{\jpsi})$ is the $v_2$ of the foreground, from same event unlike sign pairs, in [2.9, 3.3]~GeV/$c^2$. 
$N^{FG}(M_{\jpsi})$  is the number of counts in the $\jpsi$ mass range from mixed event pairs.
$N^{S}(M_{\jpsi})$ is the number of measured $\jpsi$, $N^S = N^{FG}- (N^{comb} + N^{rem})$, with $N^{comb}$ the contribution from the combinatorial background estimated using event mixing, and $N^{rem}$ the remaining contribution of the background (physical processes or uncertainty on the mixed background normalization) fitted with an exponential.
$f^{BG}(M)$ is fitted outside the $\jpsi$ mass to $v_2^{FG}(M)\cdot \frac{N^{FG}(M)}{N^{comb}(M)}$, and extrapolated to [2.9, 3.3]~GeV/$c^2$.

\section{Results}
\begin{figure} \vspace*{-3mm}
\hspace*{-1cm}
\centering
\includegraphics[clip=true,width=14cm]{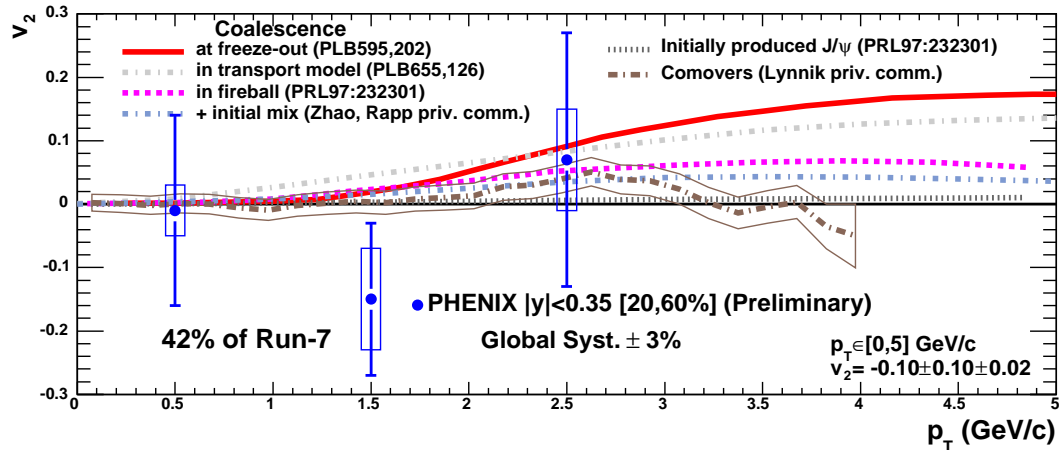}
\caption{\label{fig:v2}$\jpsi$ $v_2$ at $|y|<0.35$ for [20,60\%] in centrality, as a function of $p_T$, using 42\% of 2007 Au+Au statistics, with some predictions. Error bars are statistical errors uncorrelated point to point, boxes are systematic errors correlated between each $p_T$ bins, and a global systematic error is written (see text).}
\end{figure}

The $\jpsi$ $v_2$ at mid-rapidity for 42\% of the Run-7 statistics is shown for three $p_T$ bins in the centrality range [20,60\%] on Fig.~\ref{fig:v2}. Statistical errors and point to point uncorrelated systematic errors coming from the signal extraction are drawn with vertical bars. Systematic errors correlated between the $p_T$ bins appear with boxes. These errors account for the $v_2$ extraction and are estimated by the difference on the $v_2$ value when fitting with 
Eq.~\ref{v2 fourrier} or using directly $\langle cos2(\phi-\psi)\rangle$. They also account for the background $v_2$ shape, estimated with a fit to the background distributions with a constant, linear or quadratic polynomial. Global systematic errors to account for the technique used to determine the reaction plane angle and resolution are written on the figure.

The measurements shown on Fig.~\ref{fig:v2} have big uncertainties mostly due to lack of statistics and to the poor signal/background ratio. The $v_2$ integrated over all $p_T$ is compatible with 0: $v_2=-0.10\pm0.10\pm0.02(\pm3$\%). Some theoretical scenarios appear on Fig.~\ref{fig:v2} to illustrate different production mechanisms predictions. When only initially produced $\jpsi$ (no regeneration) are considered~\cite{Yan_init} with normal suppression from inelastic collisions with spectators and anomalous suppression from collisions with gluons in a QGP, the prediction is slightly positive but reaches at most 0.02 for $p_T=5~GeV/c$. Models with partial or full charm quark or charmonium thermalization should  have larger $v_2$ than this. The addition of recombination of $c$ and $\bar c$ during or at the term of the partonic phase should enhance the charmonium production and affect the elliptic flow. For instance, full regeneration of the $\jpsi$ at freeze out is studied in \cite{Rapp_reco_freeze_out} and in a transport model description in \cite{Ravagli_transport_model}. Those models predict a steep increase of the $\jpsi$ $v_2$, reaching 0.2 at $p_T=5$~GeV/c. If the production is a mixture of regeneration and initial production as in \cite{Zhuang_reco_cont_init,Zhao_reco_direct}, the predicted $v_2$ has a more moderate slope and reaches intermediate values. Finally, comover interactions~\cite{Linnyk_comovers} leave the $\jpsi$s with no $v_2$ within errors, slightly positive at $p_T=2.5$~GeV/c, and negative at $p_T=4$~GeV/c. The mid-rapidity $\jpsi$ $v_2$ measurement obtained with less than half of the potential 2007 statistics does not allow to distinguish between predictions. Work is ongoing to process the remaining statistics and to reduce the systematic uncertainties at mid and forward rapidities. Future RHIC Au+Au runs will provide more statistics which is crucial for this measurement. The experimental study of $\jpsi$ $v_2$ will be complementary to other studies of $\jpsi$ suppression and will provide valuable information about the early stage of high-energy heavy-ion collisions.

\end{document}